\documentclass[aps,prl,amsmath,twocolumn,longbibliography,amssymb,showpacs,floatfix]{revtex4-1}
\usepackage[dvipsnames,rgb,dvips]{xcolor}
\usepackage{graphicx,overpic,tikz}
\usepackage{siunitx}
\usepackage{bbold}
\usepackage{mathrsfs}
\newcommand{\ve}[1]{\ensuremath{\mbox{\boldmath$#1$}}}

\newcommand{\Rey}{\text{Re}_{\rm p}}

\begin{document}
\title{Collisions of micron-sized, charged  water droplets in still air}
\author{G. Magnusson$^1$, A. Dubey$^1$, R. Kearney$^2$,  G. P. Bewley$^2$, and B. Mehlig$^1$}
\affiliation{
\mbox{}$^1$Department of Physics, Gothenburg University, 
41296 Gothenburg, Sweden \\
\mbox{}$^2$Sibley School of Mechanical and Aerospace Engineering, 
Cornell University, USA 
}

\begin{abstract}
We investigate the effect of electrical charge on collisions of hydrodynamically interacting, micron-sized water droplets settling through quiescent air.  The relative dynamics of  charged droplets is determined by hydrodynamic interactions, particle and fluid inertia, and electrostatic forces.  We analyse the resulting relative dynamics of oppositely charged droplets by determining its fixed points and their stable and unstable manifolds. The stable manifold of a saddle point forms a separatrix that separates colliding trajectories from those that do not collide. The qualitative conclusions from this theory are in excellent agreement with experiments.  
\end{abstract} \maketitle

Collision-driven growth of small water droplets in air is of great importance in atmospheric physics \cite{Pru10}. Cloud
droplets tend to carry charges, in thunderclouds but also in warm rain clouds \cite{Har15}, and the resulting electrostatic forces affect droplet-collision rates and thus the droplet-size distribution.
The collision dynamics of charged cloud droplets has been studied by numerical integration of model equations. But it is not known when these models work, how they fail, how Coulomb forces compete 
with hydrodynamic effects to determine the rate at which charged droplets settling through a cloud grow by collision. There is no theory  explaining the parametric dependence  upon charge, droplet size, and velocity. 

\begin{figure}
\begin{overpic}{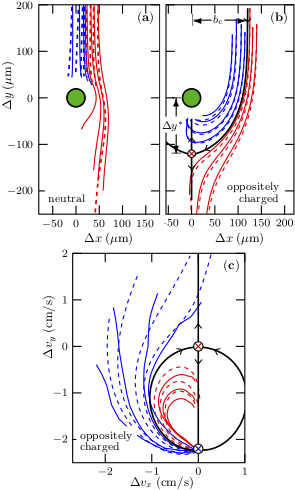}
\end{overpic}
\caption{\label{fig:RelativePhaseSpace} 
({\bf a}) Spatial separation of neutral droplets in the rest frame of the smaller droplet (green disk). 
Experiments (solid lines) and fitted simulations of Eqs.~\eqref{eq:phase_space} (dashed lines).
Blue lines indicate that the experiment resulted in a collision, red lines a miss. The inset shows false-colour snapshots from the video recording of a no-collision trajectory. 
({\bf b}) Same but for oppositely charged droplets. The inset shows a colliding trajectory.  
The red encircled cross indicates the location of a saddle point (see text) at $(0,\Delta y^\ast)$, together with its stable and unstable manifolds (solid black lines). As $t\to-\infty$, the stable
manifold connects to an unstable fixed point at $(\infty,b_c)$. 
({\bf c}) Relative velocities for charged droplets.  The blue encircled cross indicates the fixed point at infinity. 
}
\end{figure}

Even in the absence of charge, several mechanisms affect droplet collisions. Hydrodynamic interactions bend the path of one droplet around a second one [Fig.~\ref{fig:RelativePhaseSpace}({\bf a})],  reducing their chance of colliding. Particle inertia may cause  droplets of different sizes to collide nevertheless,
because it allows them to detach from the streamlines of the flow \cite{Gus16}.  However, this mechanism does not allow equally-sized droplets to approach in the creeping-flow limit, a consequence of time-reversal symmetry. Fluid inertia breaks this  symmetry and enables droplets of the same size to approach, as shown by \citet{Kle73}. However, their approximation of the hydrodynamic interactions fails as the droplets approach. Early experiments \cite{Abb74} designed to check the theory of \citet{Kle73} are inconclusive because they used an updraft to keep the droplets within the field of vision of the cameras, with unclear consequences \cite{Sar75commentA,Sar75commentB}. More fundamentally, \citet{Kle73} did not consider how the hydrodynamic approximation breaks down
below the mean free path. This is important because this effect regularises an artificial singularity of the hydrodynamic force  \cite{Kim:2005,Sun96},
which would otherwise prevent droplets from colliding.

Coulomb interactions compete with hydrodynamic effects, particle inertia, and fluid inertia.
The effects of charge on droplet collisions  is understood on a rudimentary level:
opposite charges increase collision efficiencies, and charges are more important for smaller droplets than for larger ones \cite{Pru10,Har15}. 
But there is no agreement on the number of elementary charges needed to affect the droplet dynamics.
\citet{Dav64} finds that more than 800 elementary charges are needed for the Coulomb force to have an effect for $20\,\mu$m droplets, 
 the typical size of cloud droplets, 
but \citet{Tin14} state that at least 10,000 elementary charges are required.
 \citet{Sem66} quote a limit of 6000 elementary charges. 

Early model calculations could not resolve these discrepancies because they neglected hydrodynamic interactions and fluid inertia entirely \cite{Paluch70}, causing the model to break down 
at realistic charges \cite{Abb75}. More recent model calculations \cite{Sch76,Sch79} contained hydrodynamic interactions, but only approximately, 
as in Ref.~\cite{Kle73}, and it is not known whether the model works, at least qualitatively, or not. In addition, these numerical studies were restricted to selected droplet radii and charges, and could therefore not resolve the 
uncertainties mentioned above. More importantly, lack of experimental data makes it impossible to validate the models, even for droplet collisions in still air.
Still less is known about the effects of electric fields and inhomogeneous or turbulent background flow.

We therefore investigated the collision dynamics of water droplets settling in still air, using 
Lagrangian tracking \cite{Kea20} to analyse experimental data on collisions of charged water droplets in still air. 
The data were collected with a setup similar to that described in \citet{Kelken}. 
We explain the observed collision dynamics [Fig.~\ref{fig:RelativePhaseSpace}({\bf b},{\bf c})]  by characterising  its fixed points
and their invariant manifolds. Our analysis describes how the collision 
dynamics depends on droplet charges and radii, amongst other parameters. 
The theory is accurate for the charges we consider. We discuss how it breaks down for smaller charges. 
  
{\em Background.}
Fig.~\ref{fig:RelativePhaseSpace} shows two components of the spatial separation between  two settling droplets,
$\Delta x$ and $\Delta y$,  in the rest frame of the smaller droplet.  Gravity points downwards. 
Whether the droplets collide is determined by the impact parameter $b$, the $\Delta x$-coordinate at $\Delta y = \infty$. 
The collision efficiency is defined as $b_c^2/(a_1+a_2)^2$ \cite{Pru10}, where $b_c$ is the impact parameter of the grazing path, and
$a_1$ and $a_2$ are the droplet radii.

The motion of a single water droplet in still air is controlled by three non-dimensional numbers: St, $\Rey$, and Sl. 
The Stokes number ${\rm St} \!= \!v/(a \gamma)$ determines how droplets detach from fluid stream lines due to their inertia. Here $v$ is the relative speed between the droplet and the air, 
$a$ is the droplet radius, and $\gamma = 9 \rho_{\rm f} \nu/(2 a^2 \rho_{ \rm p})$ is the Stokes constant with units of inverse time, where the fluid-mass density is $\rho_{\rm f}$, the kinematic viscosity is $\nu$, and the droplet-mass density is $\rho_{\rm p}$. 
Fluid inertia resists the motion of the surrounding air that moves along with the droplet. This effect is quantified by the particle Reynolds number $\Rey= av/\nu $. 
An accelerating droplet accelerates the surrounding air. The magnitude of this unsteady effect 
is quantified by the Strouhal number, ${\rm Sl} = a/(v \tau_{\rm c} )$, 
where $\tau_{\rm c}$ is the characteristic acceleration timescale.

In the presence of a second droplet, a fourth non-dimensional number $a/R$ (center-to-center distance $R$ between the droplets)
determines the strength of hydrodynamic interactions between the droplets, which can be computed as a perturbation expansion in $a/R$. 
In the creeping-flow limit, this corresponds to the method of reflections 
\cite{Kim:2005}.
As mentioned above, fluid inertia becomes important for similar-sized droplets. One 
can account for this effect by perturbation theory in $\Rey$  \cite{candelier2016settling}. 
A fifth parameter becomes important 
when the interfacial distance between the droplets is comparable to the mean free path, 
so that the breakdown of hydrodynamics must be considered \cite{Sun96}, 
at least for neutral droplets.

The Coulomb number, Cu, is commonly used to quantify the effect of electrical charges on the droplet dynamics.
This non-dimensional parameter  is defined as ${\rm Cu} = |E_{\rm C}|/E_{\rm kin}$, where $E_C= k_{e} q_1 q_2/(a_1+a_2)$ is the Coulomb energy upon impact
(Coulomb constant  $k_{e}$, droplet charges $q_1$ and $q_2$), 
and $E_{\rm kin}$ is the kinetic energy based on the initial relative velocity.
Defined in this way, Cu parametrises the critical collision efficiency for the central-force problem \cite{Gol02}, without dissipation or hydrodynamics interactions. 
The definition is analogous to the non-dimensional number used   to quantify the effect of electric charge on coagulation in Brownian suspensions 
\cite{hidy1965remarks}.
Lu {\em et al.}~\cite{lu2010clustering,lu2010,lu2015charged} defined an analogous parameter 
to describe the effect of charge on spatial clustering of droplets in turbulence.

\emph{Experimental methods}. 
The measurements reported here were taken using a setup similar to the one described by \citet{Kelken}. For details, see the supplemental material (SM) 
\cite{SM}. 
Droplets between $17$ and $25\,\mu$m in radius were generated using inkjet-printer nozzles \cite{sergeyev2006inexpensive}. 
The oppositely charged droplets carried about $10^5$ elementary charges. 
The neutral droplets had less than $2500$ elementary charges per droplet. 
Two nozzles were angled so that collisions and near-collisions could be observed simultaneously by two high-speed cameras. 
Electrical charge was imparted by applying voltages to the nozzles. 
The resulting charge-per-mass ratio was measured separately by allowing single droplets to fall between two parallel charged conducting plates.
We tracked the droplets 
separately in each camera plane,
using the method described by \citet{Kea20}.  We
matched droplet paths in the two cameras based on the droplet sizes, 
reconstructed the three-dimensional paths, and recorded collisions 
(for the small droplets 
considered here, all collisions led to coalescence). 
Details are given in the SM \cite{SM}. 

\emph{Model}. 
Our model takes into account particle, convective, and unsteady fluid inertia to order $a/R$. 
In the experiments, droplet Stokes numbers  were quite large,  of the order $\mathscr{O}(10)$. 
The particle Reynolds numbers were small, $\mathscr{O}(0.1)$, allowing for perturbation theory in $\Rey$  \cite{candelier2016settling}. 
The Strouhal number was $\mathscr{O} (0.1)$. Unsteady effects enter the equation of motion at order $\mathcal{O}(\sqrt{\rm Sl \, \Rey})$ \cite{Lovalenti93}, we include them by
expanding the two-droplet history force \cite{Ardekani2006} in $a/R$. Coulomb interactions are modelled by the Coulomb force
$\ve F_{e} =  k_{e} q_1 q_2 {\ve R}/R^3$, with centre-to-centre distance vector $ \ve R$ and $R=|\ve R|$.
 When the droplets are close,  induced charges give rise to corrections  calculable as an expansion in $a/R$, considering the droplets as conducting spheres \cite{Har15}.  We neglect this effect in a first approximation. The equation of motion for droplet $j$ with mass $m_j$ is
\begin{align}
 \label{eq:phase_space}
 \dot{\ve x}^{(j)}&\!=\!\ve v^{(j)} \,, \,\,\,
 \dot{\ve v}^{(j)}\!= \! \ve g + \frac{1}{m_j} ( \ve F^{(j)}_h + \ve F^{(j)}_e) \,, \,\,\,\mbox{$j\!=\!1,2$,}
\end{align}
where  $\ve g$ denotes the gravity vector.  Expressions for the hydrodynamic force $ \ve F_h$ are given in the SM \cite{SM}.

\emph{Theory}.
Relative droplet dynamics takes place in nine-dimensional space, spanned by $\ve v^{(1)}$, and relative velocity $\Delta \ve v = {\ve v}^{(2)}\!-\!{\ve v}^{(1)}$:
\begin{align} 
\dot{\ve R}&\!=\! \Delta\ve v \,, \,
 \Delta \dot{\ve v}\!= \! \frac{m_1\!+\!m_2}{m_1m_2} \frac{k_e q_1 q_2}{R^3}\ve R +\!\frac{\ve F^{(2)}_h}{m_2}\!-\!\frac{\ve F^{(1)}_h}{m_1} \label{eq:reduced_phase_space_2}  \,.
\end{align}
To characterise the droplet dynamics, we searched
for fixed points of this dynamical system and analysed their stability. We found
a saddle point where the  
droplets fall together at a fixed separation. 
For oppositely charged droplets, 
the larger droplet encounters the saddle point below the smaller one [red crosses in Fig.~\ref{fig:RelativePhaseSpace}({\bf b,\,c})]. 
The stable manifold of the saddle point connects to an unstable fixed point at infinity,  where the droplets fall at their independent settling velocities [blue cross in Fig.~\ref{fig:RelativePhaseSpace}({\bf c})]. There is a continuum of such unstable fixed points, but only one connects to the saddle point.  The connecting invariant manifold between the fixed points forms a separatrix that separates trajectories with qualitatively different behaviours \cite{Strogatz2000}.

{\em Results.}  Consider first the dynamics of neutral droplets. Fig.~\ref{fig:RelativePhaseSpace}({\bf a}) compares measured spatial separations with numerical model trajectories.
To account for experimental uncertainty, we fitted the droplet radii. The values obtained 
 are consistent with the experimental parameters within error estimates (Table~S2 \cite{SM}).
We see that the model works reasonably well, 
although
the fits become worse for separations smaller than $2(a_1\!+\!a_2)$. 
Non-colliding trajectories fit better than colliding ones. However, the non-colliding trajectory passing closest to the smaller droplet shows the largest deviation. While the experimental trajectory shows the larger droplet travelling at a fixed separation as it tumbles around the smaller one, the numerical model trajectory  collides. The model fails because because higher-order corrections in $a/R$ (lubrication forces \cite{Kim:2005}) matter at small separations. 
Even with lubrication forces though, the model cannot be used to compute collision efficiencies, because they are determined by the breakdown of the hydrodynamic approximation below the mean-free path \cite{Sun96,Dhanasekaran2021}.

Relative trajectories of oppositely charged droplets are shown in 
Fig.~\ref{fig:RelativePhaseSpace}({\bf b}). 
The electrical charges change the phase-space dynamics entirely.
Now the larger droplet overtakes the smaller one before colliding
[inset in panel~({\bf b})]. 
This cannot happen for neutral droplets, for which collisions always occur when the larger droplet is above the smaller one. 
The experimental trajectories suggest that the stable manifold of a saddle point forms a separatrix  between the colliding and non-colliding trajectories.
This is confirmed by the numerically simulated trajectories, where we fitted droplet charges in addition to droplet radii. 
The obtained charges were smaller than the experimental estimates, likely due to systematic errors in the experimental charge measurements \cite{SM}.

The corresponding relative velocities are shown in Fig.~\ref{fig:RelativePhaseSpace}({\bf c}). 
We see that the relative speeds of the droplets decrease as they approach, 
 dissipated by hydrodynamic interactions. 
All relative trajectories start close to an unstable fixed point 
where the droplets are infinitely far apart. At long times they either end up colliding, or converge to a stable fixed point where the droplets settle independently.

{\em Discussion.}
Fig.~\ref{fig:RelativePhaseSpace}({\bf b}) suggests that 
the  separatrix explains droplet-collision outcomes, in stark contrast to the neutral case, where
the breakdown of continuum hydrodynamics at small interfacial distance  $s = R-(a_1+a_2)$ determines whether droplets collide   \cite{Sun96,Dhanasekaran2021}. For charged droplets,  the attractive Coulomb force dominates at small $s$ 
because it diverges as $\sim \!s^{-1} (\log s)^{-2}$  \cite{Lek12}, and it therefore determines collision outcomes. 
Non-continuum lubrication forces ($ \sim - \Delta v \log\log (1/s)$ \cite{Sun96})  
only decrease the time to collision \cite{SM}.
Therefore
the separatrix differentiates colliding trajectories from non-colliding ones. In other words, its $\Delta x$-coordinate approaches $b_c$ as $\Delta x\to\infty$.
Non-colliding trajectories converge to a stable fixed point at infinite separation.
Colliding trajectories remain outside the circular region demarcated by the separatrix. 

\begin{figure}[t]
\begin{overpic}{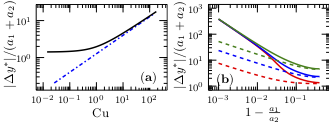}
\end{overpic}
\caption{\label{fig:saddle_point_simulations}
{(\bf a}) Location $\Delta y^\ast$ of  the fixed point in Fig.~\ref{fig:RelativePhaseSpace}({\bf b}) as a function of Coulomb number Cu,
 for Eqs.~\eqref{eq:phase_space} with $a_1/a_2=0.8$ (black), and without hydrodynamic interactions
 and without
  fluid inertia (blue, dash-dotted).  ({\bf b})  $\Delta y^\ast$ as a function of droplet-radius ratio for different charges for   the full model (solid), and excluding fluid inertia (dashed). Red, blue and green correspond to charges of $6\times 10^3, 6\times 10^4$ and $3\times 10^5$ elementary charges for one of the droplets, while the second droplets charge is kept fixed at $ 8.2 \times 10^4$ elementary charges. 
 }
\end{figure}

The saddle point emerges because gravity and hydrodynamic forces in  Eq.~\eqref{eq:reduced_phase_space_2} break the rotational symmetry of the Coulomb  problem. In the absence of these perturbations, the solutions are periodic orbits determined by the balance between the attractive Coulomb force and the repulsive centrifugal force \cite{Gol02}. 
The Coulomb number Cu determines the radius of these orbits, equal to ${\rm Cu} \, (a_1+a_2)$. In the presence of gravity and hydrodynamic effects, by contrast, the  Coulomb force is balanced by hydrodynamic forces. Solving  $\dot{\ve v}^{(1)}= \dot{\ve R}=\Delta \dot{\ve v}=0$ 
for the fixed-point  yields $(0, \Delta y^\ast, 0, v_y^\ast)$, with
\begin{equation}
|\Delta y^\ast|^2 = \frac{-k_e q_1 q_2(m_1+m_2)}{m_1 F^{(2)}_{h,y}{(\Delta y^\ast,v_y^\ast)}-m_2 F^{(1)}_{h,y}{(\Delta y^\ast,v_y^\ast)}}\label{eq:saddle_location}
\end{equation}
 (details are given in the SM \cite{SM}).
Eq.~(\ref{eq:saddle_location}) implies that $|\Delta y^\ast|\propto \sqrt{\rm Cu}$ as Cu$\to\infty$ (keeping $a_1/a_2$ fixed),
because $F^{(j)}_{h,y}$ becomes independent of $\Delta y^\ast$ in this limit.
This explains the model-simulation results shown in Fig.~\ref{fig:saddle_point_simulations}({\bf a}).  
It also explains why $ \Delta y^\ast$ becomes independent of hydrodynamic interactions and fluid inertia at large Cu
[Fig.~\ref{fig:saddle_point_simulations}({\bf a})]:
the  corresponding terms  in $F_{h,y}^{(1,2)}$  vanish as {$|\Delta y^\ast| \to\infty$} (SM \cite{SM}). 

The situation is quite different for similarly-sized droplets. 
 Fig.~\ref{fig:saddle_point_simulations}({\bf b}) 
indicates that fluid inertia causes $|\Delta y^\ast|$ to become insensitive to charge as $a_1 \to a_2$, and that fluid inertia plays a major role in this.
When the droplets are far apart, Coulomb attraction ($\propto R^{-2}$) is negligible compared to attraction by fluid inertia ($\propto R^{-1}$).
Solving for  $\Delta y^\ast$, we find $|\Delta y^\ast |\sim |a_2-a_1|^{-1}$ as $a_1 \to a_2$ (details in the SM \cite{SM}).
As $|\Delta y^\ast|$ diverges, so does $b_c$:  droplets always collide in this limit.

Finally, the linear stability matrix at the saddle point determines the time $t_{\rm esc}$ a droplet takes to pass close to the saddle point  
\cite{Strogatz2000}, e.g. to reach a distance of $a_1+a_2$ from it. This  time
diverges 
as $t_{\rm esc} \sim -\tfrac{1}{\lambda_+}\ln |b-b_c|$, where  $\lambda_+$ is the leading eigenvalue of the stability matrix. Numerical diagonalisation of the stability matrix of Eqs.~\eqref{eq:reduced_phase_space_2} (but ignoring the history force) gives $\lambda_+^{-1} \approx 3.46$, in good agreement with the 
value of  $t_{\rm esc}$ obtained from integrating model trajectories \cite{SM}.

{\em Conclusions.} 
We explained the qualitative nature of the collision dynamics of oppositely charged droplets settling in still air, and determined its parameter dependence
by analysing its fixed points, their stability, and their invariant manifolds. We found that  a manifold connecting 
a saddle and an unstable fixed point   separates between colliding trajectories and non-colliding ones. Our analysis predicts 
how Coulomb forces compete with inertial and hydrodynamic forces to determine the collision dynamics. The main conclusions of our analysis are in good agreement with the experimentally observed collision dynamics of small charged droplets settling in still air, and with  the  results of our model simulations.

The experimental data clearly suggests that there is a saddle point for oppositely charged droplets [Fig.~\ref{fig:RelativePhaseSpace}({\bf b})], yet there is no such saddle point for neutral droplets [Fig.~\ref{fig:RelativePhaseSpace}({\bf a})]. How does the saddle point form as the droplets become charged? A bifurcation at a critical Coulomb number could explain  how many elementary charges are needed to qualitatively change the collision dynamics. 
We therefore expect that dynamical-systems analysis as described in this Letter, analysing the collision dynamics in terms of its fixed points and their bifurcations, will resolve this question. 
Since collision dynamics is discontinuous, bifurcations mechanisms observed in non-smooth dynamical systems \cite{Hogan} may matter here.

At small Cu, our model predicts that the saddle point occurs when the droplets are close. In this limit, higher-order corrections in $a/R$ must become important: not only to the electrostatic force (caused by induced charges), but also to the hydrodynamic force (lubrication effects). For similar-sized droplets one must account for 
fluid inertia, but it is a challenge to find reliable approximations  when the droplets are close. At small charges, the precise location of the separatrix
is surely affected by how non-continuum effects regularise lubrication forces. 

Concerning droplet growth by collision and coalescence, note that collision kernels depend not only upon the critical impact parameter but also on relative velocities \cite{Gus14c}. Since this velocity vanishes at the saddle point, droplets may collide at smaller relative velocities compared to neutral droplets, at least at small charges where the saddle point occurs for closeby droplets. This could give weakly charged droplets 
more time to interact.  

While our charges are considerably smaller  than those in the innovative study of spatial droplet patterns in turbulence~\cite{lu2010clustering,lu2010,lu2015charged}, they are nevertheless by about one order of magnitude 
 larger than in warm  clouds \cite{Tak73}. 
 To validate models for the dynamics of closely approaching droplets, measurements for weakly charged droplets are necessary.
 
Updrafts in clouds cause  air turbulence which affects collisions of droplets settling through the cloud. How do turbulent fluctuations change the phase-space dynamics, how do they interplay with the mechanisms described above? Little is known, but recent progress was 
made in describing collision rates of neutral droplets in straining flow; numerical model simulations exhibit a rich variety of different dynamical behaviours, as well as intricate parameter dependencies \cite{Dhanasekaran2021}. The underlying phase-space dynamics is probably quite different from that
shown in Fig.~\ref{fig:RelativePhaseSpace}({\bf b},{\bf c}), but it is clear that a systematic bifurcation analysis will reveal the underlying mechanisms.

A next step towards understanding the effect of flow-gradients and ultimately turbulence is to analyse general steady linear flows (e.g. rotational flows 
\cite{Brunk98}). 
A long-term goal is to understand the effect of cloud-droplet charges on fluctuations of spatial separations and relative velocities of droplets in turbulence. How does the spatial clustering observed by Lu {\em et al.}~\cite{lu2010clustering,lu2010,lu2015charged} develop as the charges become smaller? This is a challenging question, not least because recent experiments \cite{Yav18,Bragg21} suggest evidence for extreme spatial clustering of small droplets in turbulence. Which mechanism causes this remarkable effect is an entirely open question.


%

\end{document}


\title{Supplemental Information for \lq{}Collisions of micron-sized, charged water droplets in still air\rq{}}

\author{G. Magnusson$^1$, A. Dubey$^1$, R. Kearney$^2$,   G. P. Bewley$^2$, and B. Mehlig$^1$}
\affiliation{
\mbox{}$^1$Department of Physics, Gothenburg University, 41296 Gothenburg, Sweden \\
\mbox{}$^2$Sibley School of Mechanical and Aerospace Engineering, Cornell University, USA}
\maketitle
\tableofcontents

\section{Experimental set up}
\label{sec:set up}

\begin{figure}[b]
\begin{overpic}[width= 10cm]{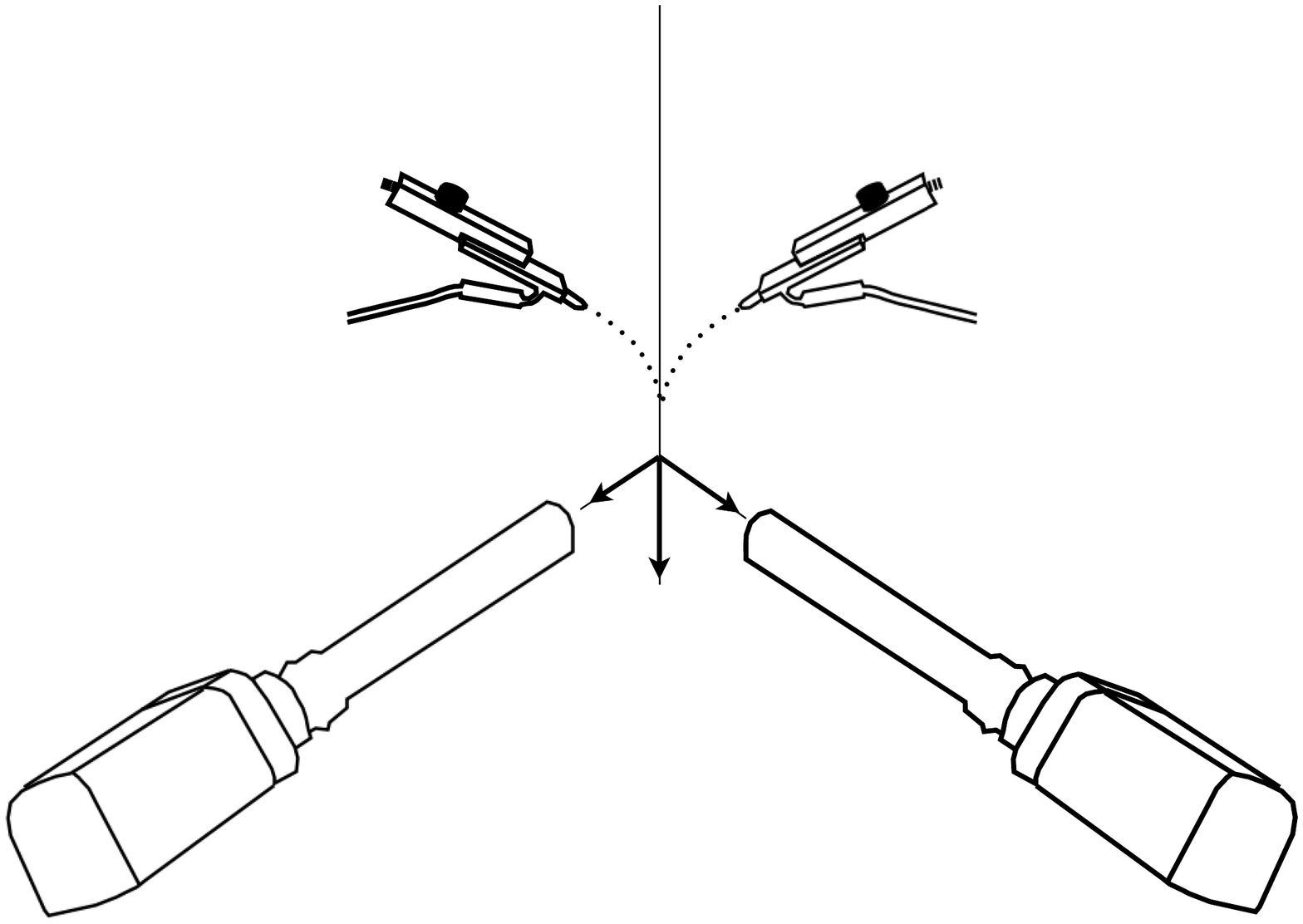}
\put(47,22){$-\hat{\bf e}_y$}
\put(42,32){$\hat{\bf e}_z$}
\put(55,31){$\hat{\bf e}_x$}
\end{overpic}
\caption{\label{fig:schematic} 
Experimental set up. 
Shown are the two droplet dispensers (top), the two cameras at right angles (bottom), and two droplets paths. 
The axes of the cameras are parallel with the x- and z-axes.}
\end{figure}

We analysed collisions between pairs of droplets approaching each other in still air using a set up similar to \citet{Kelken} and \cite{Ivanov2017}. 
Droplets were generated using inkjet-printer technology as described by \citet{sergeyev2006inexpensive}.
Two printer nozzles were angled at each other such that collisions and near-collisions occurred in the observation volume where the fields of view of two high-speed cameras (Phantom Miro Lab310 from Vision Research) were aligned.
Optical traps \cite{Kelken} were not used.
We defined the lab-frame coordinates to be collinear with the axes of the cameras, so that one camera pointed along the $z$-axis (viewing images in the $x$-$y$ plane), and the other was aligned with the $x$-axis (viewing images in the $y$-$z$ plane).
Gravity pointed in the negative $y$-direction to within $3$ degrees. 
The setup is shown schematically in Figure \ref{fig:schematic}. The cameras were equipped with long distance microscopes (Model K2 DistaMax from Infinity) so that the droplet images were magnified by about $11$ times.
The resolution of each camera was measured by taking a picture of a flat calibration mask (Thorlabs model R2L2S3P1).
The spatial resolution of each camera was was found to be $1.78$ and $1.70 \mu$m/pixel.
The uncertainty in the resolution is $\pm 10\%$ \cite{Ivanov2017}.

The droplet generators were triggered such that pairs of droplets were generated one after another. 
This allowed between $1$ and $40$ approaching droplet pairs to be recorded in the same measurement before the memory of the cameras filled up. 
The droplet pairs were generated slowly enough that no more than two droplets were in the observation volume at once.
Charge was imparted to the droplets by applying voltage to an electrode near the generator.
Each measurement was stored in the Phantom camera .cine format, which includes both the image data and metadata such as date of collection and frame rate. 
The data files are labelled as follows: {\texttt{YYYY\_MM\_DD\_measure\_XX\_collision\_XX}}. 
Here,  {\texttt{YYYY\_MM\_DD}} refers to the date on which the measurement was performed. The numbers following {\texttt{measure}} and  {\texttt{collision}} are labels for different measurements. 
The sampling rate of the cameras were set to either $8,500$ or $25,000$ frames per second, depending on the experiment.
 
A total of $1291$ droplet pairs were observed, with radii between $20$ and  $26 \mu$m and impact parameters  $b$ between about $0.1$ and $6$.
As explained in the main text, the impact parameter is the component of the separation distance between the droplets that is orthogonal to the relative velocity between them; 
in the absence of external forces, it is equal to $0$ when the droplets approach each other head-on, and equal to $1$ during a grazing collision where the droplets touch edges with the relative velocity perpendicular to the separation vector.
The impact parameter varied from event to event because of randomness in the droplet generation process, and it varied between different sets of experimental conditions because of changes to the droplet generators' positions and angles.
The initial velocity of the droplets could be controlled by changing the signal sent to the droplet generators, but this signal was not altered during the experiments.
In this supplemental information, we describe all the data collected during this experiment, though not all data is used in the main paper.
Two video recordings of representative near-contact events are also available with this supplemental material.
 
\section{Data analysis}
\subsection{Tracking}
\label{sec:Tracking}
 
We estimated the radii and positions of droplets in each camera plane using the circle-fitting method of Pratt \cite{Pratt1987} which accurately finds the sub-pixel center location of overlapping circles in digital images \cite{Kea20}. 
\citet{Kea20} tested this method on synthetic digital images that have similar intensity profiles to the present experimental data, and demonstrated that the algorithm measures the sizes and positions of droplets in high-resolution 
more precisely than other methods, in particular when the droplet images overlap in the camera plane.

Because the internal clocks of the two cameras were not synchronized, we tracked the droplets separately in each camera plane using the hybrid method discussed in Ref.~\cite{Kea20}.
We then found the time delay between the cameras  using the moment of collision as a reference.
We do not consider data from any video recordings in which no collisions took place.
We reconstructed the three-dimensional trajectories assuming the optical axes of the cameras were orthogonal, taking the $x$- and $y$-coordinates from one camera and the $z$-coordinates from the other one.

For all analyses, droplet radii were measured using only images from the camera with its axis colllinear with the z-axis (hereafter referred to as camera 1) because the images recorded by the other camera (camera 2) were not uniformly illuminated, causing a downward bias in the particle sizes measured by the algorithm in regions of less intense lighting.
We found that the measured droplet radius was near its minimum in the time signal along a trajectory when the image was in best focus, so we used the 10\textsuperscript{th} percentile of the signal as our estimate for the true droplet size.
We used the 10\textsuperscript{th} percentile rather than the minimum because it is robust to excursions due to noise \cite{Einarsson2013}.

Droplet velocities were calculated using a two-point, forward finite-difference method and smoothed with a Gaussian filter of window size of about $1$ or $0.4$ ms, depending on the frame rate of the cameras.
This filter size is large enough that it reduces the effect of noise due to uncertainty in the droplet positions, but it is smaller than the relaxation time of the droplets.
We discarded short trajectories (fewer than $10$ data points) that were more likely to contain tracking errors. 

The uncertainty in finding the droplet positions in the digital images is less than half a pixel \cite{Kea20} or about $0.9 \mu$m.
Because the optical transfer function that maps the measured droplet sizes to their true sizes is not known in detail, we used information near collision to help estimate the droplet sizes.
We used as an upper bound for the sum of the radii the minimum separation distance immediately prior to collision plus the uncertainty in finding the droplet positions, mentioned above.
We used as a lower bound the predicted separation distance in one time step (using the velocity estimates at the end of each trajectory) minus the uncertainty in finding the droplet positions.
We found good agreement between the mean of these two bounds and the radii of the droplets measured directly from camera 1 once a correction factor of $0.85$ was applied.
A correction factor less than one indicates that the particle identification algorithm overestimates the droplet sizes, which is consistent with a Gaussian blur (used during particle identification) and the diffraction pattern arising from from an opaque sphere both increasing the apparent size of the images.

\subsection{Collision detection}

The collision detection method discussed in Ref.~\cite{Kea20} cannot be applied to find collisions in this data because the time it takes two parent droplets to coalesce into one daughter droplet was longer than the time between each image recorded by the cameras.
Instead, in order to determine collision outcome, we used the following criteria:
\begin{enumerate}
    \item Two trajectories end (parents) and a new trajectory begins (daughter) within at most $16$ frames.
    \item	The daughter must have a radius equal to the calculated radius that preserves the mass of the parents (within threshold $2\,\mu$m).
    \item The daughter may not coexist with any other trajectories.
\end{enumerate}
The first criterion provides a permissive threshold for the coalescence time, the time it takes the daughter droplet to achieve a stable spherical shape (rather than oscillating between a prolate and oblate spheroid).
Because the Weber number of the droplets was small, there was no fragmentation upon collision, so the second criterion enforces the physical requirement of conservation of mass.
The final criterion prevents a simultaneous interruption in two trajectories from causing a false positive collision detection.
If all these criteria are fulfilled, we considered a collision to have occurred.

If a collision is not found by the criteria above, the event could either be labeled as a miss (non-collision) or inconclusive. 
If the inward component of the relative velocity vector was positive at the final moment of observation, then we marked the event as inconclusive.
These are cases in which the collision might or might not have occurred outside the observation volume of the cameras.
Otherwise, the event was marked as a miss.

\subsection{Charge Measurements}
\label{sec:ChargeMeasurements}

As described in Ref.~\cite{Kelken}, the charge-per-mass ratio was measured before and after data collection by allowing single droplets to fall between two parallel capacitor plates, where each trajectory was observed by a web-camera with a large field of view so that the full path of the droplet between the plates could be observed.
The linear size of the plates was much larger than that of the region of observation, so it can be assumed that the electric field was homogeneous.
The voltage on the charging electrode was changed manually until the desired deflection of the droplet path was achieved for each of the two dispensers.
To determine the direction of gravity, the single droplets were first observed falling without any voltage applied. 

\begin{figure}
   \includegraphics{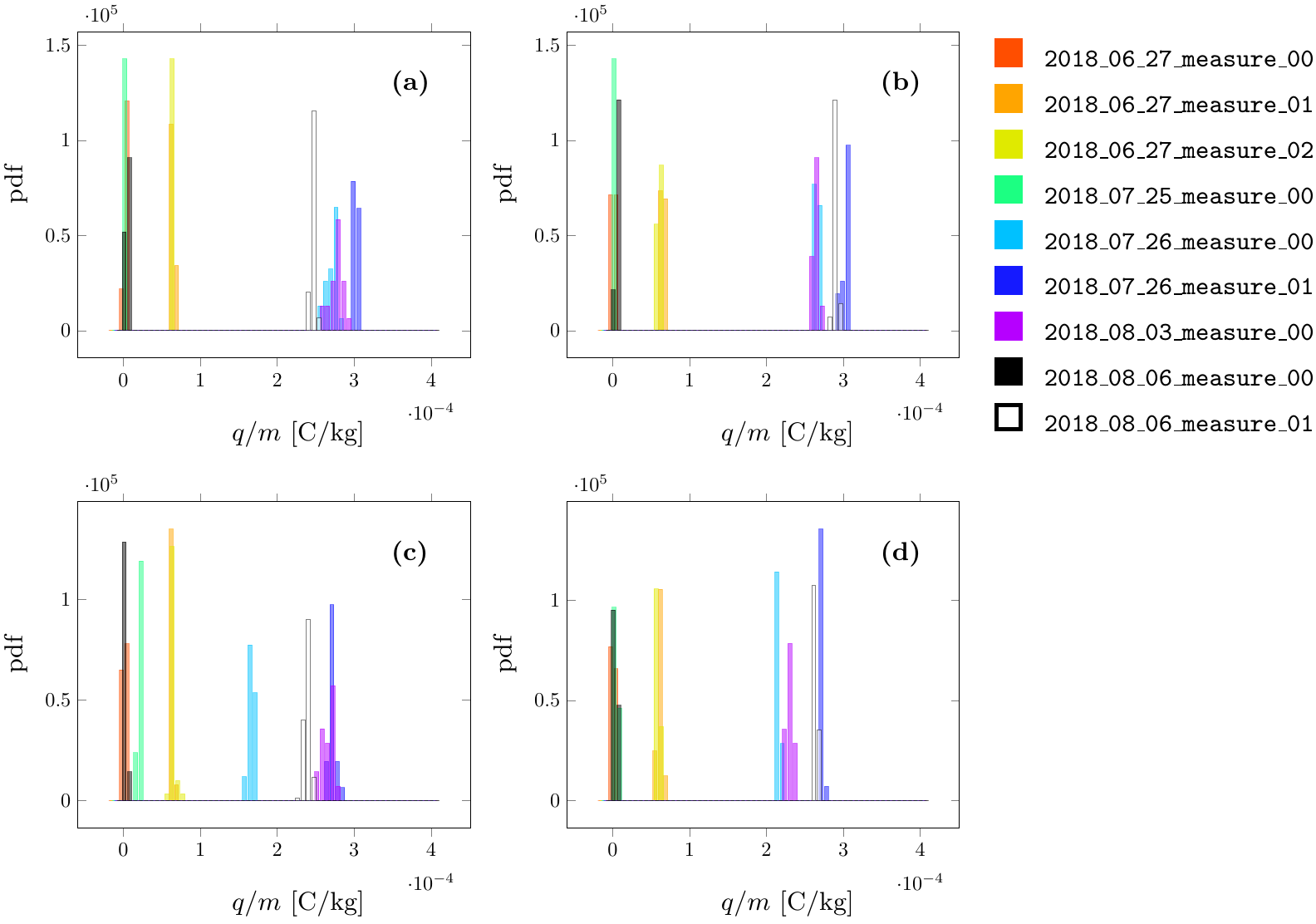}
    \caption{   
    Estimated droplet charge-per-mass ratios.
    (\textbf{a}): Primary droplet, before the data were collected.
    (\textbf{b}): Satellite droplet, before the data were collected.
    (\textbf{c}): Primary droplet, after the data were collected.
    (\textbf{d}): Satellite droplet, after the data were collected.
    The color of each distribution indicates the name of the measurement according to the legend.}    
    \label{fig:qm}
\end{figure}

The deflection angle $\alpha$ of the settling path due to charge on the plates was used to calculate the charge-per-mass by
\begin{equation}
    \label{eq:qm}
    \frac{q}{m}=\frac{g d}{U}\tan \alpha
\end{equation}
where $g$ is the acceleration due to gravity, $d$ is the separation distance between the plates ($d=17$ mm), and $U$ is the applied voltage difference across the plates (which was set to $100 V$). 
The distribution of $q/m$  estimated from Eq.~(\ref{eq:qm}) for each set of data is shown in Figure \ref{fig:qm}.
We refer to the smaller droplet in each pair as the primary droplet and the larger droplet as the satellite droplet.
The charge-to-mass ratio varies between  $1 \times 10^{-6}$ and $3 \times 10^{-3}$ C/kg.

The variation in charge-to-mass ratio before and after data collection for each measurement was within $5 \times 10^{-5}$ C/kg except for the measurement \texttt{2018\_07\_26\_measure\_00} wherein it varied by $1 \times 10^{-4}$ C/kg for the primary droplet.
The large shift in the distribution of charge-to-mass ratio in Figure \ref{fig:qm} from before to after data collection was most likely due to misalignment between the axis of the web-camera and the direction of gravity; 
an upward or downward tilt of the camera would tend to cause an overestimate of the deflection angle.
Additionally, it was necessary to clean the dispenser nozzles at regular intervals to prevent clogging, which may have altered the charge imparted to the droplets.

\section{Range of parameters studied}
\label{sec:range}

\begin{figure}
\includegraphics{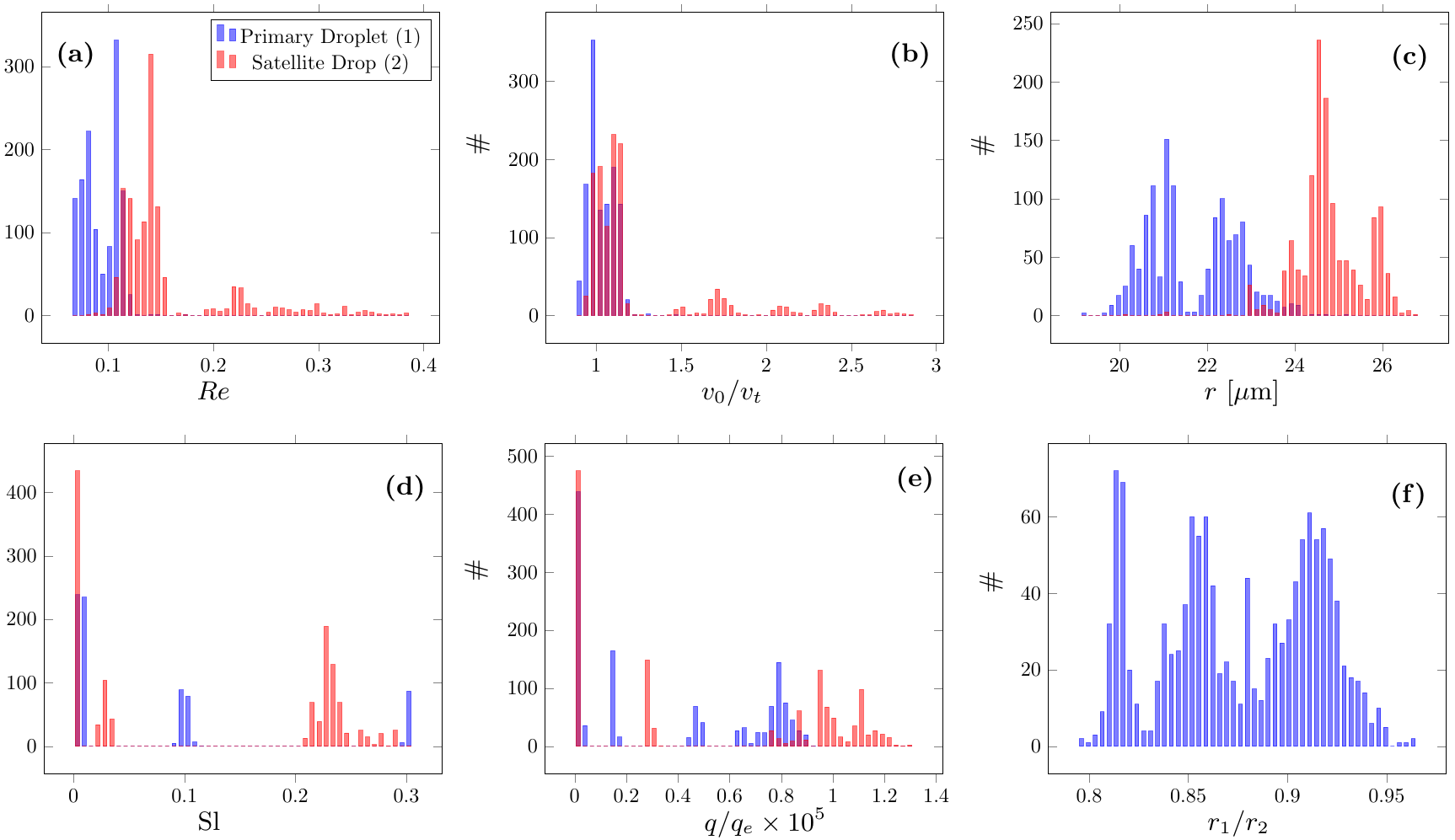}    \caption{
    ({\bf a}) Distribution of droplet-Reynolds numbers.
    ({\bf b}) Distribution of initial speeds as a ratio of the Stokes terminal velocity.
    ({\bf c}) Distribution of droplet radii.
    ({\bf d}) Distribution of the droplet-Strouhal number.
    ({\bf e}) Distribution of droplet charges in number of electrons.
    ({\bf f}) Distribution of radius ratio between primary and satellite droplet.
    }
    \label{fig:DropletCharacteristics}
\end{figure}

Figure \ref{fig:DropletCharacteristics} shows a summary of some important droplet characteristics. 
Droplet Reynolds numbers are shown in Figure \ref{fig:DropletCharacteristics}({\bf a}).
We define $\Rey = \frac{a v}{\nu}$ as in the main text, where $a$ is the droplet radius, $v$ is the average speed of the droplet over the entire time it is observed, and $\nu = 1.48 \times 10^{-5}$ m$^2$/s is the kinematic viscosity of air. 
The data where the satellite droplet achieved $\Rey > 0.2$ are from \texttt{2018\_06\_20} and \texttt{2018\_06\_27}.
In these events, the satellite droplets had initial velocities that were large compared to their Stokes settling speed because they retained momentum from their emission from the droplet generator.
In later measurements, the droplet generators were adjusted so the satellite droplets had more time to relax to their settling speed.
The droplet Reynolds numbers were small but nonzero.

The ratio of the initial speed of the droplets $v_0$ to their Stokes settling speed $v_g$ are shown in Figure \ref{fig:DropletCharacteristics}({\bf b}).
The settling speed is calculated using $v_g = \frac{2}{9} \frac{\rho_p}{\nu \rho_f} a^2 g$, 
where $\rho_p = 1000$ kg/m$^3$ is the mass density of water, 
$\rho_f = 1.23$ kg/m$^3$ is the mass density of air,
and $g = 9.81 m/s^2$ is the acceleration due to gravity.
The initial speed of the droplets was of the same magnitude as the Stokes settling speed except for measurements performed on \texttt{2018\_06\_20} and \texttt{2018\_06\_27}, where, as mentioned above, the satellite droplets had initial velocities that were large because they retained momentum from their emission from the droplet generator. 

The measured droplet radii are shown in Figure \ref{fig:DropletCharacteristics}({\bf c}).
An estimate for the droplet radius was measured at each time the cameras observed a droplet.
As discussed in Section \ref{sec:Tracking}, we used only size estimates from camera 1 and applied a correction factor of $0.85$ to map from the radius of the shadows cast by the droplets to their actual radii.
The values shown here are the distribution of the 10\textsuperscript{th} percentile from each droplet trajectory.
The uncertainty in measuring the droplet radius from the particle identification algorithm is no more than half a pixel, which corresponds to about $0.9 \mu$m.
The dominant source of uncertainty in measuring the radii comes from uncertainty in the spatial resolution, which is $\pm 10\%$.

The distribution of the Strouhal number, Sl = $\tfrac{a}{v \tau_c}$ is shown in panel ({\bf d}). Here $\tau_c = \sqrt{\tfrac{a m (a_1+a_2)^2}{k_e |q_1 q_2|}}$ is a timescale based on droplet accelerations due to Coulomb forces, for charged droplets with $m$ the mass of the droplet. The distributions show that the Strouhal number ranges up to 0.3.

The droplet charges are shown in panel ({\bf e}).
We estimated the charges using the average charge-per-mass for each measurement obtained from the calibration procedure (results shown in Figure \ref{fig:qm}) and the mass of the droplets measured from their size assuming the density was $1000$ kg/m$^3$.
The droplets in the present experiment carried between several hundred to about $10^5$ elementary charges, either positive or negative. 
The oppositely charged droplets, discussed in Figure 1 of the main text, carried between $0.6 \times 10^5$ to $1.2 \times 10^5$ elementary charges. 
For comparison, 25 $\mu$m radius droplets in thunderstorm clouds carry an average charge of about $0.2 \times 10^5$ elementary charges while warm clouds can contain droplets with an average charge of about $0.2 \times 10^4$ elementary charges \cite{Tak73}.
Most other studies considered much larger charges. 
\citet{Abb75} studied droplets droplets settling in still air with $1 \times 10^6$ to $8 \times 10^6$ elementary charges (near $1 \times 10^{-5}$ C/$m^2$). 
Lu {\em et al.} \cite{lu2010clustering,lu2010,lu2015charged} measured how charges affect spatial clustering of particles in turbulence. 
Their droplets carried from $3 \times 10^5$ to $4 \times 10^5$ elementary charges.
The maximum charge a droplet can contain before electrostatic forces overwhelm surface tension and the droplet bursts is given by $q^2 = 64 \pi^2 \epsilon_0 \gamma a^3$ \cite{Rayleigh},
where $q$ is the maximum charge the droplet can sustain,
$\epsilon_0$ is the permittivity of free space,
and $\gamma$ is the surface tension of the droplet.
For a droplet with a radius of $20 \mu$m, for example, the maximum charge it can contain is about $10^7$ elementary charges. 
The charges in this experiment are well below this limit.

Uncertainty in estimating the charge on each droplet arises from uncertainty in the mass of the droplets, which is calculated from the measured radii, and from uncertainty in the charge-per-mass measured from the calibration procedure.
For measurements with approximately uncharged droplets, uncertainty from the calibration procedure dominates; 
these droplets may contain between $0$ and several thousand electrons.
For measurements with highly charged droplets (excluding \texttt{2018\_07\_26\_measure\_00}, discussed below), uncertainty in the mass of the droplets dominates; 
these droplets may contain $\pm 30\%$ the number of electrons indicated in Figure \ref{fig:DropletCharacteristics}({\bf e}).
As discussed in Section \ref{sec:ChargeMeasurements}, the uncertainty in measuring the charge-per-mass of the droplets in the measurement \texttt{2018\_07\_26\_measure\_00} was substantially larger, so both uncertainty in the mass of the droplets and uncertainty in the charge-per-mass play a role in the total uncertainty.
For this measurement only, the uncertainty in charge contained on the droplets is $\pm 56\%$ for the primary droplet and $\pm 37\%$ for the satellite droplet.

The distribution of ratio of primary droplet radius to satellite droplet radius is shown in Figure \ref{fig:DropletCharacteristics}({\bf f}).

\section{Equations of motion}
In this Section we give the details of the model used in our analysis (Eq. (3) in the main text \cite{maintext}). The hydrodynamic force is computed as an expansion in $a/R$ using the method of reflection \cite{Kim:2005}. The method of reflection gives an iterative solution of the Oseen equation (valid when the Reynolds number is small, as in our case, see Section \ref{sec:range}) using its Green function \cite{candelier2016settling}. The fluid flow close to the droplet is approximated by creeping flow, and the two solutions are matched to obtain a uniformly valid solution. The result is that the velocity induced by a given droplet at a position $\ve R$ in a coordinate frame with the origin at the droplet center is given by \cite{candelier2016settling},

\begin{equation} \label{eq:fluid_velocity}
 \ve u =  e^{-\frac{1}{2 \nu} (v R + \textbf{\emph{v}} \cdot \ve R)} \frac{\textbf{ \emph f}}{8 \pi \mu R} +\left\lbrace 1-\left(1 +\frac{v R}{2 \nu} \right) e^{-\frac{1}{2 \nu} (vR + \textbf{\emph{v}} \cdot \ve R)} \right\rbrace \frac{ f}{{\rho}_{\rm f}  v} \frac{\ve R}{4 \pi R^3}.
\end{equation}
Here $\ve v$ is the droplet velocity relative to the fluid, $\nu$ is the fluid viscosity, $R =|\ve R|$, and $\rm{\rho}_f$  is the fluid density. The force $\ve f$ is the force applied by the droplet on the fluid and is given by $\ve f = 6 \pi \mu a \ve v$, where $a$ is the droplet radius. The force acting on the first droplet due to the second droplet is given by,

\begin{equation} \label{eq:f_hydro}
 \ve F^{(1)} = - 6 \pi \mu a_1 (1 + \Rey_{1}) (\ve v^{(1)} -\ve u^{(2,1)}).
\end{equation}
Here, $\ve u^{(2,1)}$ is the fluid velocity induced at the position of droplet 1 due to droplet 2 calculated using Eq.~\eqref{eq:fluid_velocity}.  Eq.~\eqref{eq:f_hydro} gives an approximation to the hydrodynamic force, valid when the inter-droplet distance is much larger than their radii. This approximation fails at close separations, and higher orders in $a/R$ must be considered. The models used in previous studies, for instance by \citet{Kle73} and \citet{Sch76}, used similar approximations which failed when the droplets separation became small. \citet{Kle73}, for instance,  used an ad-hoc modification of the Oseen equation to compute the effect of fluid inertia on small droplet collisions, assuming that the droplet dynamics is two-dimensional. They performed an ad-hoc rescaling of the Reynolds number, $\Rey \to \Rey[1-0.08 \log (1+50 \Rey)]$ \cite{Car53, Pru10} to account for overestimation of the fluid inertia effects in Oseen equations. In addition, neither our model nor the models in the studies mentioned above account for continuum breakdown when the interfacial droplet separation is of the order of the mean free path of air, and thus cannot describe collisions. In order to accurately model droplet collisions, first, fluid inertia must be included to higher orders in $a/R$. Most importantly, lubrication effects must be described, which determine forces on the droplets when their interfacial separation is smaller than either droplets radius. Second, breakdown of the continuum approximation must be included because this describes the inter-droplet forces when their interfacial separation is of the order of the mean-free path of air \cite{Sun96}.

Electrical forces accelerate the droplets as they approach one another. In the experiments, the Strouhal number is $\sim 0.1$, the same order as the Reynolds number, see Section \ref{sec:range}. Since the unsteady effect enters the force at order $\mathcal{O}(\sqrt{{\rm Sl} \, \Rey})$ \cite{Lovalenti93},  history forces must be taken into account.  We use the expressions derived by \citet{Ardekani2006}, who obtained the history force acting on a droplet in the presence of a second droplet, reproduced in their notation:

\begin{align} \label{eq:f_history}
 \ve F_{\text {history}}^{(1)}= - 6 \pi \mu a_1 \int_{0}^t d\tau \, \left\lbrace \frac{d \ve v^{(1)} (\tau)_\perp }{d \tau} \tilde{g}_1^*(t-\tau)   +\frac{d \ve v^{(1)} (\tau)_\parallel}{d \tau} \tilde{h}_1^*(t-\tau)
 -\frac{d \ve v^{(2)} (\tau)_\perp}{d \tau} \tilde{g}_2^*(t-\tau)
 -\frac{d \ve v^{(2)} (\tau)_\parallel}{d \tau} \tilde{h}_2^*(t-\tau) \right\rbrace.
\end{align}
Here, the subscripts $\perp$ and $\parallel$ denote components of velocity perpendicular and parallel, respectively, to the droplet separation vector. Their results are valid to order $(a_1/R)^3$, but, in order to be consistent with the rest of our hydrodynamic treatment, we only keep term to order $ \epsilon =a_1/R$ by computing the asymptotics of the expressions computed by \citet{Ardekani2006} in the $\epsilon \to 0$ limit. The result is
\begin{subequations}
\begin{align}
      g_1(t) &=     \begin{cases}
       \frac{a_1}{ \sqrt{\pi \nu \, t}}&\quad\text{if} \quad t\ll \frac{a_1^2}{\nu \epsilon^2}  \,;\\
       \frac{a_1}{ \sqrt{\pi \nu \, t}} (2- 9 \beta \epsilon) &\quad\text{if} \quad t\gg \frac{a_1^2}{\nu \epsilon^2} \,,
       \end{cases} \\
       h_1(t) &=    \frac{a_1}{ \sqrt{\pi \nu \, t}},\\
       g_2(t) &=  \begin{cases}
       -\frac{3}{4} \beta \epsilon &\quad\text{if} \quad t\ll \frac{16 a_1^2}{9 \pi \nu \epsilon^2}  \,;\\
       \frac{a_1 \beta (-4 +3 (1+\beta)\epsilon)}{4 \sqrt{\pi \nu \, t} } &\quad\text{if} \quad t\gg \frac{16 a_1^2}{9 \pi \nu \epsilon^2} \,,
       \end{cases} \\
        h_2(t) &=  \begin{cases}
       -\frac{3}{2} \beta \epsilon &\quad\text{if} \quad t\ll \frac{4 a_1^2}{9 \pi \nu \epsilon^2}  \,;\\
       \frac{a_1 \beta (-2 +3 (1+\beta)\epsilon)}{2 \sqrt{\pi \nu \, t} } &\quad\text{if} \quad t\gg \frac{4 a_1^2}{9 \pi \nu \epsilon^2} \,.
       \end{cases}
\end{align}
\end{subequations}
Here, $\beta =a_2/a_1$. The time integral in Eq.~\eqref{eq:f_history} starts at time $0$ instead of at $-\infty$. An extra term must be added to that equation if the fluid velocity does not match the initial droplet velocity \cite{Michaelides92}. Since we expect that initially the droplet acceleration with respect to the fluid is negligible, we have ignored this extra term as a first approximation. The total hydrodynamic force on droplet 1 is computed as the sum of the forces in Eq.\eqref{eq:f_hydro} and Eq.\eqref{eq:f_history}. The force on the second droplet can be similarly obtained by interchanging the indices $1 \leftrightarrow 2$ in Eqs.~\eqref{eq:f_hydro} \textendash~\eqref{eq:f_history}.

\begin{figure}[ht!]
       \includegraphics{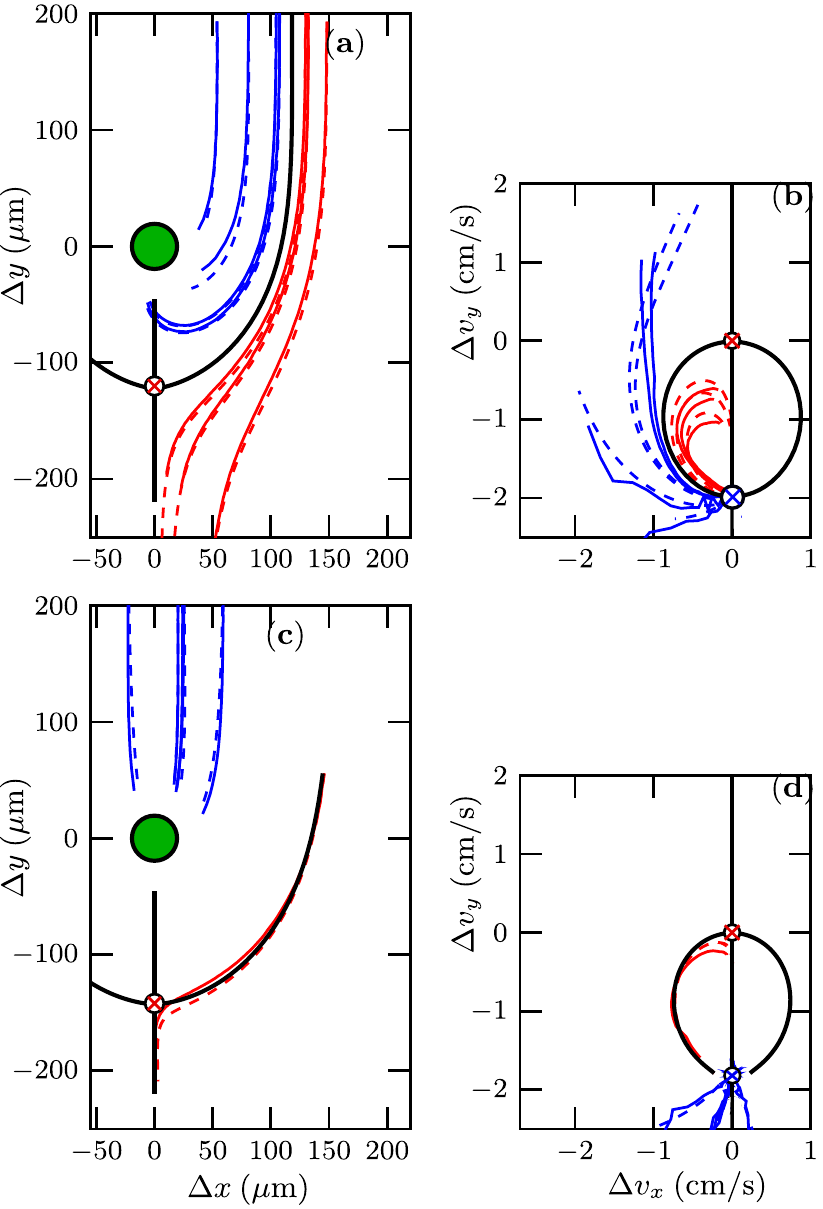}
\caption{\label{fig:RelativePhaseSpace} 
({\bf a,c}) Relative separation of oppositely charged droplets in the frame of reference of the smaller droplet (green disk at the origin). 
Experiments (solid lines) and fits to simulations of  Eqs.(2) in the main text \cite{maintext} (dashed lines).
Blue lines indicate that the experiment resulted in a collision, red lines a miss. The red encircled cross indicates the location of a saddle point (see text) together with its stable and unstable manifolds (solid black lines).  
({\bf b,d}) Relative velocities for charged droplets.  The blue encircled cross indicates the second fixed point. ({\bf a,b}) show trajectories in the measurement {\texttt{2018\_07\_26\_measure\_00\_collision\_02}}, whereas ({\bf c,d}) show measurement {\texttt{2018\_07\_26\_measure\_00\_collision\_00}}.
}
\end{figure}
\section{Fixed points and their stability}
The dynamical system consisting of the nine-dimensional space spanned by $\ve v^{(1)},\ve R$ and $\Delta \ve v$ with equations of motion given by Equations (2) and (3) in the main text \cite{maintext} exhibits fixed points: a saddle point where the larger droplet travels below the smaller one, and a continuum of fixed points at infinite separation, unstable for $\Delta y = +\infty$ but stable for $\Delta y=-\infty$. In this Section we examine these fixed points and their stability. The saddle point is found by solving the following equations for the fixed point,
\begin{subequations}
\begin{align}
 \dot {\ve v }^{(1)} &= 0\,, \label{eq:saddle_1}\\
 \dot {\ve R}\,\,\, &= 0 \, \label{eq:saddle_2},\\
 \Delta \dot  {\ve v } \,\,\,&= 0\, \label{eq:saddle_3}.
\end{align}
\end{subequations}
The first equation implies that the velocity of the first droplet is constant at the saddle point. The third equation constrains the droplet relative velocity at the saddle point to remain constant, the second equation says that this relative velocity vanishes. Eqs.~\eqref{eq:saddle_1} -- \eqref{eq:saddle_3} lead to two implicit equations for the common settling velocity of the droplets, $\ve v^\ast = (0,v_y^\ast,0)$ and the droplet separation at the saddle point $ {\ve R}^\ast = (0,\Delta y^\ast,0)$,
\begin{align}
  \ve g + \frac{1}{m_1} ( \ve F^{(1)}_h + \ve F^{(1)}_e) &=0 \, ,\\
 \frac{m_1+m_2}{m_1 m_2} \frac{k_e q_1 q_2}{R^3} \ve R +\frac{\ve F^{(2)}_h}{m_2}-\frac{\ve F^{(1)}_h}{m_1} &=0 \,.
\end{align}
Using the hydrodynamic force derived in the previous Section, the final equations can be written as,
\begin{align}
  g -\frac{k_e}{m_1} \frac{q_1 q_2}{ \Delta y^2 }- 6 \pi \mu a_1 (1+\frac{3 a_1 v_y^\ast}{8 \nu}) \frac{v_y^\ast}{m_1}+ 9 \pi \mu a_1 a_2 (1+\frac{3 a_1 v_y^\ast}{8 \nu}) \frac{v_y^\ast}{m_1}\frac{1}{\Delta y} &=0 \, , \label{eq:saddle_final1}\\
g+ \frac{k_e}{m_2} \frac{q_1 q_2}{\Delta y^2} -\frac{1}{m_2} 6 \pi \mu a_2 (1+\frac{3 a_2 v_y^\ast}{8 \nu}) \left[v_y^\ast+ \left( {\rm e}^{-\tfrac{\Delta y ^\ast v_y^\ast}{\nu}}-1\right)\frac{3}{2}\frac{a_1 \nu}{\Delta y}\right]&=0 \,.\label{eq:saddle_final2}
\end{align}
Note that the history force does not appear in the above equations because the droplets experience no acceleration at the saddle point. Equations~\eqref{eq:saddle_final1} and \eqref{eq:saddle_final2} must be solved for $v_y^\ast$ and $\Delta y^\ast$. Asymptotic expressions for $\Delta y^\ast$ can be obtained in the limit of large charge and when $a_1 \to a_2$. As the charge magnitude increases, so does $\Delta y^\ast$, while $v_y^\ast$ asymptotes to a constant value. Consequently, at large charges the first three terms in Eq.~\eqref{eq:saddle_final1} are of the same order while the last term vanishes. Similarly, the second term in square brackets in Eq.~\eqref{eq:saddle_final2} vanishes as well.  These equations can be solved to obtain the asymptotic dependence $\Delta y^\ast \propto \sqrt{q_1 q_2} \propto \sqrt{\rm Cu}$, for fixed $a_1$ and $a_2$. In the limit $a_1 \to a_2$, $\Delta y^\ast$ diverges as well. In this limit, one can ignore the Coulomb terms in equations~\eqref{eq:saddle_final1} and \eqref{eq:saddle_final2}, as well as the exponential term in Eq.~\eqref{eq:saddle_final2}. This results in an asymptotic dependence $\Delta y^\ast \sim (a_2-a_1)^{-1}$. These arguments demonstrate the claims in the discussions Section of the main text regarding Figure 2 \cite{maintext}.

\begin{figure}[t]

\begin{overpic}[scale=1.0]{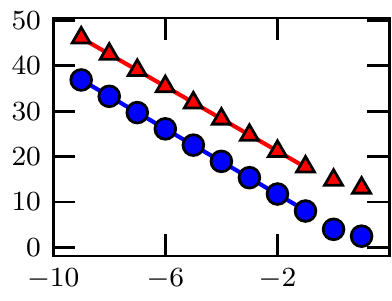}
\put(30,-6){{$\log|b-b_c|$}}
\put(-7,30){\rotatebox{90}{ $t_{\rm esc}/\tau_0$}}
\put(42,60){{slope $=-3.58$}}
\end{overpic}
\caption{\label{fig:saddle_point_simulations}
 Escape time $t_{\rm esc}$ normalised by the interaction timescale $\tau_0= (a_1+a_2)/|\Delta \ve v_0|$ plotted as a function of the difference of the impact parameter, $b$, to the critical impact parameter, $b_c$, defined in the main text \cite{maintext}. 
Blue circles correspond to data points for which $b < b_c$, red triangles are for $b>b_c$. Red and blue lines show fitted exponential laws with slope $-3.58 \pm 0.02$.  
 }
\end{figure}

The stability of this saddle point (ignoring the history force) can be determined by calculating the eigenvalues, $\lambda_i, i=1,\dots, 9$, of the Jacobian at the saddle point. Numerical diagonalisation gives one eigenvalue with a positive real part, and and eight eigenvalues with negative real parts, ${\rm Real}(\lambda_1) > 0 > {\rm Real}(\lambda_2) \geq {\rm Real}(\lambda_3) \geq \dots \geq {\rm Real}(\lambda_9)  $. The eigenvalue with positive real part corresponds to an eigenvector directed along the positive $\Delta y$ and $\Delta v_y$ directions. This describes how trajectories escape along the $y-$axis. Numerical computation for typical experimental parameters [$a_2 = 19.5\mu$m, $a_1 = 23.7\mu$m, $q_1 = 6.5 \times 10^{-15}$C (41,000 elementary charges), and $q_2 = -1.2 \times 10^{-14}$C (74,000 elementary charges)] gives $ {\rm Real}(\lambda_1) = 139.9/$second. Normalizing this by the interaction time $\tau_{0}= \tfrac{a_1+a_2}{|\Delta v_{0}|}$ based on the initial relative velocity of the droplets gives $\lambda_+^{-1}= 3.46$. The positive eigenvalue gives the scaling exponent of the time to escape from a saddle point \cite{Strogatz2000}. Our result is in good agreement with simulations of the escape time from the saddle point, shown in Fig.~\ref{fig:saddle_point_simulations}. The Figure demonstrates that simulations of the escape time, including the history force, give a scaling $ t_{\rm esc} \sim -3.56 \log |b-b_c|$ in good agreement with numerical computation of the real part of the positive eigenvalue $= 3.46$. We expect that the reason for the good agreement is that the history force is a small correction, which does not drastically modify escape time.

Next we describe the dynamics close to the fixed points at infinity. When the droplets are infinitely far apart, they settle at their respective settling velocities. There is a fixed point of the relative droplet dynamics when $\Delta y = \pm \infty$. But how do the dynamics change when the droplets are at a large but finite separation? At a large separation, $ \Delta \dot y  = \Delta v_y < 0$ in our convention. Thus, $\Delta y$ decreases from a large separation so that if $\Delta y >0$ the system moves away from the fixed point but if $\Delta y<0$ the system moves towards the fixed point at infinite separation. Thus, the fixed points at infinite separation are unstable when $\Delta y > 0$, but stable when $\Delta y <0$.

\section{Hydrodynamic and electrostatic forces at small separations}

In the absence of charges, hydrodynamic lubrication forces between two spheres prevents them from colliding. For neutral droplets it is therefore important to account for continuum breakdown at small separations. \citet{Sun96} showed that this allows droplets to collide in finite time. Their argument was based on the fact that  hydrodynamic dissipation diverges as the particles approach, and their initial kinetic energy will be completely dissipated before the particles can collide. They further presented a simple model which could be analytically solved and demonstrated that the relative velocity between spheres vanishes before they can collide. 

In the case of charged droplets, however, the model cannot be analytically solved and it is not known whether the droplets can collide. In this section, we show that charged droplets can collide in finite time despite a repulsive hydrodynamic singularity, due to a competing singularity in the attractive electrical force. We further argue that the weaker singularity of the non-continuum force \cite{Sun96} speeds up the collision process by decreasing the time until collision.

The electrostatic force between two conducting, charged spheres can be calculated as series expansion in $a/R$. When the interfacial separation, $s=R-(a_1+a_2)$, between the spheres is smaller than the two radii, the force may be efficiently expressed as a series in $a/s$, which diverges as $s \to 0$, where the leading term (for two equally sized spheres) is $F_e = - k_e (q_1-q_2)^2 [2 a s \, (\log \tfrac{4a}{s})^2]^{-1}$ \cite{Lek12}. In order to understand the dynamics of charged, hydrodynamically interacting droplets at close approach, we consider a model similar to \citet{Sun96} (but ignoring continuum breakdown) with the additional electrical force, $F_e$. The equation of motion now becomes,
\begin{align}
\dot s &= \Delta v,\label{eq:s}\\
 \Delta \dot v &= - 6 \pi \mu a^2 \frac{\Delta v}{ s} - \frac{k_e (q_1-q_2)^2 }{ 2 a s (\log \tfrac{4a}{s})^2} \label{eq:dv}.
\end{align}
In the limit $s\to 0$, the equation Eq.~\eqref{eq:dv} becomes overdamped and the relative velocity must obey,
\begin{align}
 \Delta v = -\frac{k_e (q_1-q_2)^2 }{12 \pi \mu a^3} \frac{1}{(\log \tfrac{4a}{s})^2}.
\end{align}
Solving Eq.~\eqref{eq:s} gives the time until collision, which remains finite. For typical parameter values in the experiment considered, this timescale is of the order of $10^{-5}$ seconds, with the initial interfacial separation $s_0 = 0.1 a$. Thus, we have shown that the divergence of the electrical force for small interfacial separations allows droplets to collide despite a repulsive hydrodynamic lubrication force. The analysis is performed for droplets with the same size, but the result holds for droplets with radius ratio close to unity as well. Non-continuum corrections to the lubrication force exhibit a weaker singularity $\sim -\Delta v \log \log \tfrac{1}{s}$ \cite{Sun96}. Thus, collisions of charged droplets would occur sooner compared to the continuum case when taking non-continuum effects into account.  
\section{Fitting of experimental data}
\begin{table}
\begin{center}
\begin{tabular}{ |c | c| c| } 
\hline
\multirow{2}{*}{ Measurement} &\multirow{2}{*}{ Radius Fitted({\bf Reported})  [$\mu$m]}& \multirow{2}{*}{Charge Fitted ({\bf Reported}) [$\tfrac{q_1 q_2}{e^2}10^{-9}$]} \\
 & & \\
 \hline
\verb|2018_07_25_measure_00_collision_08| &  & ({\bf 0.0037$\pm$0.003})\\
\obs{494} & \multirow{8}{*}{\shortstack[1]{20.5 ({\bf 21.3$\pm$2.0})\\24.4 ({\bf 25.2$\pm$2.5})}} & -\\
\obs{487} & &-\\
\obs{489} & &-\\
\ans{475} & &-\\
\ans{483} & &-\\
\ans{480} & &-\\
\ans{492} & &-\\
\obs{497} & &-\\
\hline
\verb|2018_07_26_measure_00_collision_03| &  & ({\bf-4.1$\pm$2.7})\\
\obs{390} & \multirow{8}{*}{\shortstack[1]{19.5 ({\bf 20.0$\pm$2.0})\\23.7 ({\bf 24.2$\pm$2.4})}} & -3.4388\\
\obs{392} & &-3.4652\\
\obs{393} & &-3.3997\\
\ans{395} & &-3.4035\\
\ans{402} & &-3.3148 \\
\ans{394} & &-3.3662\\
\ans{386} & &-3.0987\\
\ans{403} & &-3.4289\\
 \hline
 \verb|2018_07_26_measure_00_collision_04| &  & ({\bf-4.0$\pm$2.6})\\
  \obs{347} & \multirow{8}{*}{\shortstack[1]{20.4 ({\bf 20.6$\pm$2.1})\\ 24.4 ({\bf 24.0$\pm$2.4})}} & -2.5983\\
\obs{352} & &-2.6106\\
{357} & &-2.4186\\
\ans{362} & &-2.4534  \\
\ans{367} & &-2.2638\\
\ans{372} & & -2.2688\\
{377} & & -2.4077\\
 \hline
 \verb|2018_07_26_measure_00_collision_02| &  &({\bf-4.1$\pm$2.8}) \\
 \ans{405} & \multirow{8}{*}{\shortstack[1]{20.0 ({\bf 20.3$\pm$2.0})\\ 23.7 ({\bf 24.0$\pm$2.4})}} & -2.8858\\
\obs{408} & &-3.1250\\
\ans{411} & &-2.9430\\
\obs{414} & &-2.9892\\
\ans{417} & &-2.7313\\
\obs{420} & & -2.8954\\
\obs{423} & & -2.9454\\
\ans{426} & & -2.2981\\
\ans{406} & & -2.9431\\
 \hline
 \verb|2018_07_26_measure_00_collision_00| &  & ({\bf-3.6$\pm$2.4}) \\
 \ans{446} & \multirow{6}{*}{\shortstack[1]{20.3 ({\bf 20.3$\pm$2.0})\\ 23.4 ({\bf 23.0$\pm$2.3})}} & -3.5093\\
\ans{447} & &-3.3692\\
\obs{449} & &-3.3110\\
\ans{450} & &-3.4071\\
\ans{451} & &-3.2807\\
\ans{452} & &-3.5480\\
 \hline
 \verb|2018_08_06_measure_01_collision_02| &  & ({\bf -9.7$\pm$4.2}) \\
 \ans{1042} & \multirow{6}{*}{\shortstack[1]{24.1 ({\bf 23.8$\pm$2.4})\\ 26.1 ({\bf 25.0$\pm$2.5})}} & -5.8879\\
\obs{1043} & &-6.2777\\
\ans{1044} & &-6.0872\\
{1048} & &-6.3287  \\
\ans{1049} & &-5.6761\\
\ans{1050} & &-5.7901\\
 \hline
\end{tabular}
\end{center}
\caption{\label{tab:table2} Table comparing fitted charges and radii to experimentally reported values. The first column shows the measurements and the numbered events within each measurement (arbitrary indexing for events). Blue corresponds to a colliding trajectory while red to a non-colliding one. Black corresponds to an inconclusive trajectory (see Section III). The second column shows the fitted droplet radii with the experimentally reported radii bracketed in bold text. The third column shows the squares of fitted charges (in units of $10^{-9}$ elementary charges squared), with the corresponding reported values bracketed in bold text. The measurements \texttt{2018\_07\_25\_measure\_00\_collision\_08} and  \texttt{2018\_07\_26\_measure\_00\_collision\_03} correspond to the events plotted in Fig.~1({\bf a}) and Fig.~1({\bf b,c}), respectively,  in the main text \cite{maintext}.    }
 \end{table}

In this Section we discuss the fitting procedure used to check the consistency between the model equations of motion, and the experiments. Figure 1 in the main text \cite{maintext} shows the experimental and fitted trajectories for two measurements, one with neutral droplet pairs and one with charged droplet pairs. Figure \ref{fig:RelativePhaseSpace} in this Supplementary Material shows the relative dynamics for two measurements (with charged droplet pairs) from Table~\ref{tab:table2}. Table~\ref{tab:table2} compares the fitted radii and charges to those measured experimentally.  While the fitted radii agree well with the reported radii, the fitted charges are smaller than the experimentally reported charges, see Section \ref{sec:range} and Table \ref{tab:table2}. Note that Table~\ref{tab:table2} shows the charge squared which has an error about twice as large as the charge magnitudes. We expect this difference to be caused by systematic errors in alignment of the web-camera used to measure the fall angle of single droplets for determining charge (see Section \ref{sec:range}).  Using droplet radii and charges as fitting parameters, we used least-squares fitting to fit the droplet trajectories obtained by numerical integration of the model, Eq.~(1) in the main text \cite{maintext}, to the experimental trajectories. The initial positions and velocities were chosen to be the same as in the experiments. For each event, the experimental relative separation in time and the center of mass position in time was fitted to the corresponding trajectories obtained by numerical integration of the model equations.  The fitting parameters were chosen to be radii for the two droplets, and the product of their charges. We assumed that the droplets sizes did not vary significantly within a measurement so that there are only two radii parameters per measurement. This claim is backed up by our observations. Thus, during fitting, multiple events within the same measurement were simultaneously fitted assuming the droplet radii were the same for all events. The reason for fitting products of droplet charges is that this is the relevant parameter determining the Coulomb force. We fitted a different value of this product per event, to account for uncertainty in droplet charges, see Figure~\ref{fig:qm} which shows a shift in the distributions of the charge-per-mass even for measurements performed on the same day (see Section \ref{sec:range}). Thus, for $n$ fitted events, the number of fitting parameters were $n+2$. The trajectories were fitted from the moment when the center-to-center droplet separation was $6(a_1+a_2)$, up until they reached a separation of $2(a_1+a_2)$  for the first time, or until they were last observed together.


%